\documentclass[prd,twocolumn,showpcs,amsmath,amssymb,nofootinbib,preprintnumbers,balancelastpage,longbibliography]{revtex4-1}
\usepackage[utf8]{inputenc}
\usepackage{feynmf}
\usepackage{epsfig}
\usepackage{amsmath}
\usepackage{bm}
\usepackage{times}
\usepackage{physics}
\usepackage{graphicx}
\usepackage[table]{xcolor}
\usepackage{slashed}
\usepackage{graphicx}
\usepackage{amsmath}
\usepackage{mathrsfs}
\usepackage{tikz}
\usetikzlibrary{positioning,shapes}
\usepackage{relsize,caption} 
\usepackage{textcomp}
\usepackage[T1]{fontenc}
\usepackage{tikz}
\usepackage{amsfonts}
\usepackage{booktabs}
\usepackage{siunitx}
\graphicspath{ {images/} }
\usepackage{hyperref} 
\hypersetup{breaklinks=true, colorlinks=true, citecolor=cyan, linkcolor=blue, urlcolor=blue,filecolor=blue}

\usetikzlibrary{trees}
\usetikzlibrary{decorations.pathmorphing}
\usetikzlibrary{decorations.markings}
\usetikzlibrary{positioning,arrows}
\usetikzlibrary{decorations.pathmorphing}
\usetikzlibrary{decorations.markings}
\usetikzlibrary{decorations.pathreplacing,calc}
\usetikzlibrary{decorations.pathmorphing,decorations.markings,trees,positioning,arrows}   
\newif\ifmirrorsemicircle

\usepackage{amsfonts}
\usepackage{amsmath}
\usepackage{graphicx}
\usepackage{hyperref}
\usepackage{anyfontsize}
 \usepackage{ booktabs}
\usepackage{ulem}
\usepackage{lineno}
\usepackage{epsfig,graphics,rotate,color}
\usepackage{feynmf}
\usepackage[com pat=1.1.0]{tikz-feynman}
\usepackage{array}

\def\be{\begin{equation}}
\def\ee{\end{equation}}
\def\bea{\begin{eqnarray}}
\def\eea{\end{eqnarray}}
\def\bec{\begin{center}}
\def\enc{\end{center}}

\def\bean{\begin{equation*}}
\def\eean{\end{equation*}}

\begin{document}

\title{Diquark Induced Short-Range Nucleon-Nucleon Correlations \& the EMC Effect}

\newcommand*{\LBNL}{Nuclear Science Division, Lawrence Berkeley National Laboratory, Berkeley, CA 94720, USA}\affiliation{\LBNL}
\newcommand*{\JLAB}{EIC Center at Jefferson Laboratory, Newport News, VA 23606, USA}\affiliation{\LBNL}

\author{Jennifer~Rittenhouse~West}\email{jennifer@lbl.gov} \affiliation{\LBNL}\affiliation{\JLAB}

\date{\today}

\begin{abstract}
Diquark formation across a short-range nucleon-nucleon pair is proposed as the underlying QCD physics of short-range correlations (SRC) in nuclei.  SRC pairs have been proposed as the cause of distorted quark behavior in nuclei; experimentally observed quark momentum distribution distortions termed the EMC effect.  The strong spatial overlap of SRC pairs brings nucleon constituents within range of inter-nucleon QCD potentials and any bonds formed - such as the diquark bond - affects their distributions.   In this SRC model, diquarks form in the $\rm 3_C \otimes 3_C \rightarrow  \bar{3}_C$ channel of $\rm SU(3)_C$ acting on valence quarks from highly overlapping nucleon wavefunctions.  The most energetically favorable diquark is a valence $u$ quark from one nucleon with a valence $d$ quark from the other in a spin-0 state bound together via continual single gluon exchange and an attractive quantum chromodynamics short-range potential.  Formation of a new scalar isospin-singlet $[ud]$ diquark across a NN pair is proposed as the primary QCD-level theoretical foundation for SRC models of distorted structure functions in $\rm A\geq 3$ nuclei.  Contributions from the higher mass spin-1 isospin triplet states $(ud)$, $(uu)$ and $(dd)$ are possible, with the spin-1 $(ud)$ diquark proposed as a higher mass but viable structure function distortion mechanism for the spin-1 ground state deuteron. Predictions are made for lepton scattering experiments on $\rm ^3H$ and $\rm ^3He$ nuclear targets, with implications for the coefficients of the 3-valence quark Fock states in the nucleon wavefunction.
\vspace{11mm}
\end{abstract}

\maketitle

\section{Introduction}
\label{sec:intro}
Diquarks are quark-quark correlations formed on short-range quantum chromodynamics (QCD) length scales, $d_{qq}\lesssim 1~\rm fm$  \cite{Anselmino:1992vg,Jaffe:2004ph,Barabanov:2020jvn,Wilczek:2004im,Peskin:2019iig}.  In this work, diquark formation across two nucleons via the attractive QCD quark-quark potential is proposed as the underlying QCD-level source of short-range correlations in nuclear matter and the cause of distortions in quark behavior in the nuclear environment.

Theoretical models of altered quark behavior in the nucleus in the form of nuclear structure function modifications have been studied since the European Muon Collaboration's deep inelastic scattering (DIS) experiments at CERN first revealed the surprising distortions now known as the EMC effect in 1983 \cite{Aubert:1983xm}.  Two leading explanations for the EMC effect have emerged over time: multi-nucleon mean field models \cite{Cloet:2019mql,Weinstein:2010rt} and the 2-body nucleon-nucleon short-range correlation (SRC) model \cite{Norton:2003cb, Hen:2012fm}.  Strong evidence for short-range nucleon-nucleon correlations as a source of distortions of measured nucleon structure functions when bound in nuclei has been found by the CLAS collaboration \cite{CLAS:2019vsb}.  The existence of short-range correlated NN pairs in the nucleus has long been known \cite{Brueckner:1955zzd,1981PhR....76..215F}and SRC were first proposed as the cause of the EMC effect over 30 years ago \cite{1990PhRvC..41.1100C,CiofidegliAtti:1991ae}.  

A QCD basis for short-range nucleon-nucleon correlations in all nuclei is given in this work.  For neighboring nucleons with sufficient wavefunction overlap, $u$ and $d$ valence quarks with opposite spins are predicted to form a new low-mass scalar diquark via the attractive potential of the antisymmetric $\rm SU(3)_C$ channel $3_C \otimes 3_C \rightarrow \overline{3}_C$.  The triplet set of isospin-1 spin-1 diquarks can also contribute to this effect, in particular in the deuterium nucleus (Sec.\ref{sec:hh}), but their higher masses suppress their contributions to ground state wavefunctions.

The wavefunction of the nucleon-nucleon (NN) system with the newly formed diquark is no longer a state of two distinct color singlets but a mixed state containing a minimum of one and up to three diquarks.  Color-charged scalar diquarks may take on the role of Cooper pairs in $\rm U(1)_{\rm EM}$, potentially breaking $\rm SU(3)_C$ symmetry by their formation in a two nucleon system \cite{Rapp:1997zu}.  The maximal scalar diquark formation is the tri-diquark state $[ud][ud][ud]$, a  6-quark object that is suppressed from combining into a $J\!=\!0$ color singlet due to spin-statistics constraints on the di-nucleon system.  This is relevant for the deuteron wavefunction, allowing for spin-1 vector diquark induced structure function distortions as discussed in Sec.\ref{sec:hh}.  The deuteron is an anomalous case in the diquark formation model due to the color confinement constraint on the $n-p$ system; there is no external nuclear medium for QCD diquarks in $\rm ^2H$ to move through. 

A 12-quark color-singlet solution to the EMC effect has been published, a 6-diquark state dubbed the ``hexadiquark,'' that modifies structure functions for the $^4\rm He$ nucleus and all $A\geq4$ nuclei \cite{West:2020rlk}.  The diquark formation model presented in this work acts twice within the hexadiquark and, in addition, can occur in every NN pair, includes the higher mass isospin triplet diquarks and is offered as the basis of SRC in nuclei.  As such, diquark formation must occur in all $\rm A\!=\!3$ nuclei with SRC present.  The MARATHON experiment at Jefferson Lab \cite{Abrams:2021xum,Cocuzza:2021rfn,Segarra:2021exb} is expected to publish EMC effect results in the near future.  The E12-11-112 experiment at Jefferson Lab with  $\rm A\!=\!3$ nuclear targets has published their isospin dependent SRC results \cite{Li:2022fhh}.   Predictions for $\rm ^3He$ and $\rm ^3H$ data from the diquark formation model are given in this work.

\section{NN Diquark Formation Model}
\label{sec:formation}
Experimental and theoretical nuclear physics have shown that approximately $20\%$ of nucleons in nuclei are in short-range correlations, i.e., pairs of nucleons that have fluctuated into a high relative momentum state \cite{Hen:2016kwk,CiofidegliAtti:2015lcu}.  High relative momentum between nucleons translates into large spatial overlap which can bring quarks from distinct nucleons into QCD interaction range.  A diquark  formed in the overlap region is proposed as the cause of the correlations between nucleons, with each nucleon donating a quark to the bound state.

Diquark formation between quarks from nearest neighbor nucleons is based on the following three criteria:
\begin{itemize}
    \item Quark-quark coupling: Viability of the $3_C \otimes  3_C \rightarrow \overline{3}_C$ channel of QCD
    \item Quark-quark potential: $V(r_{\rm q-q})$   attractive and binding at short distances
    \item Diquark binding energy: Much greater than average nuclear binding energy, QCD energy scales
\end{itemize}
The first two criteria are well-known calculations applied in a novel setting, i.e. across the NN pair.  The third criterion, diquark binding energy, is calculated in Sec.\ref{sec:mass}.  Auxiliary arguments in support of  diquark formation are also included within this work, e.g. in the quark-quark separation distance estimates of Sec.\ref{sec:mass}.  

Diquark bonds, or correlations, are a QCD combination of two quarks each in the fundamental representation of $\rm SU(3)_C$ transformed into an antifundamental representation, $3_C \otimes 3_C \rightarrow \overline{3}_C$.  The quarks in a diquark are continually exchanging single gluons for the duration of the bond. 

Diquark formation utilizes the attractive potential between two quarks in the fundamental representation of $\rm SU(3)_C$, calculated to have exactly half the strength of a color singlet $3_C \otimes \overline{3}_C \rightarrow 1_C$ potential \cite{Peskin:2019iig,Peskin:1995ev},
\bea 
V(r)=-\frac{2}{3}\frac{g_s^2}{4\pi r},
\eea 
where $g_s$ is the QCD coupling, to create a quark-quark bond between a pair of nearest neighbor nucleons.  The binding energy of the diquark is calculated in Sec.\ref{sec:mass}.

Estimations of the nucleon-nucleon separation distance sufficient for diquark formation are in Sec.\ref{sec:mass}.  An upper limit distance scale for $V(r_{\rm q-q})$ to act is taken to be the separation between the centers of masses of adjacent (and in contact) nucleons.  For a proton charge radius of $r_p=0.84~ \rm fm$  \cite{cite-key,Mihovilovic:2020dmd,PhysRevC.93.055207,10.1093/ptep/ptaa104} and a neutron magnetic radius of $r_n=0.86~ \rm fm$ \cite{Tanabashi:2018oca}, this gives a rough estimate for in-contact but non-overlapping NN of $d_{\rm NN} < 1.72~ \rm fm$.

For nucleons with high enough relative momenta $\Delta p_{\rm NN}$, the distance between nearest neighbor nucleons becomes smaller than the diameter of the nucleon \cite{Hen:2016kwk}, roughly for $\Delta p_{\rm NN} \gtrsim 300~ \rm MeV/c$, a momentum value above the nuclear Fermi momentum of $k_{F} \approx 2 m_{\pi}\approx 250~ \mathrm{MeV/c}$ \cite{Duer:2018sby}.  In nuclei, quantum fluctuations in the spatial separation between neighboring nucleons have been proposed to achieve sub-nucleon diameter overlap with densities up to $4$ times larger than typical nuclear densities \cite{Sargsian_2003}.  The quark-quark $\rm SU(3)_C$ potential can act upon quarks across the individual nucleon color singlets during quantum fluctuations in NN spatial separation as well as fluctuations in NN relative momentum.

The lowest mass $[ud]$ diquark combines the $\rm SU(3)_C$ triplet up and down flavor quarks in isospin singlet and spin singlet states to form an overall antisymmetric wavefunction upon exchange of quarks,
\bea \label{dq}
 |\psi_{[ud]}\rangle ^a 
\!=\!  \frac{1}{2} \epsilon^{abc}  |d_b^{\downarrow}~ u_c^{\uparrow}\! - d_b^{\uparrow}~ u_c^{\downarrow}\! - u_b^{\downarrow}~ d_c^{\uparrow} \!+ u_b^{\uparrow}~d_c^{\downarrow}\rangle.
\label{eq:diquark}
\eea 
The spin-1 isospin-1 diquark wavefunctions are given here for completeness,
\bea 
 |\psi_{(ud)}\rangle ^a 
\!=\!  \frac{1}{\sqrt{2}} \epsilon^{abc}  |d_b^{\uparrow}~ u_c^{\uparrow}\! + d_c^{\uparrow}~ u_b^{\uparrow}\! \rangle,
\eea 
\bea 
 |\psi_{(uu)}\rangle ^a 
\!=\!   \epsilon^{abc}  |u_b^{\uparrow}~ u_c^{\uparrow}\! \rangle,
\eea 
\bea 
 |\psi_{(dd)}\rangle ^a 
\!=\!  \epsilon^{abc}  |d_b^{\uparrow}~ d_c^{\uparrow}\! \rangle.
\eea 
All diquark wavefunctions have zero orbital angular momentum between quarks and therefore symmetric S-wave spatial wavefunction components.

Diquarks with spin-0 and isospin-0, $[ud]$, are considered to be ``good" diquarks in the sense that they are lighter mass than the vector diquarks, earlier estimated to be $200 ~\rm MeV$ lighter \cite{Jaffe:2004ph} which is quite close to the value found in Sec.\ref{sec:mass}.  ``Bad'' diquarks are the higher mass spin-1 isospin-1 diquarks.  The early mass estimation was based on the masses of $\Delta$ baryons and $N$ nucleons; {\it{e.g.,}} for the  $S=\frac{3}{2}$ $\Delta^0$ baryon with quark content $(ud)d$ and the $S=\frac{1}{2}$ neutron with quark content $[ud]d$, the difference in diquark mass $ m_{(ud)} - m_{[ud]}$ was approximated by  $\frac{2}{3} \left( M_{\Delta^{0}} - M_{N^{0}}\right) \approx 200~\rm MeV$.  Square brackets denote scalar diquarks and parentheses denote the spin-1 vector diquarks.  Thus the most energetically favorable configuration of the diquark is the $[ud]$ state.  The spin-1 isospin-1 diquark $(ud)$ will also be available as a binding mechanism between nucleons, despite its higher mass, as will the $(uu)$ and $(dd)$ isospin-1 triplet companions.  Formation of any of these diquarks will contribute to the observed distortions of quark distribution function in nuclei.  Here the primary focus is on the most energetically favorable state, $[ud]$.  

In the quark-diquark model of baryons the only nucleon pairs with available $u$ and $d$ valence quarks are proton-neutron pairs.  This can be seen by comparing the $u[ud]$ proton quark-diquark content  with the $d[ud]$ quark-diquark content of the neutron.  Two protons can only form a higher mass $(uu)$ diquark between them since the other valence quarks are already bound in $[ud]$ diquarks.  Similarly for two neutrons which can only form a $(dd)$ diquark between them.  Only a proton-neutron pair can form the low mass $[ud]$ because their free valence quarks are $u$ and $d$, respectively.  In this scenario, diquark induced short-range correlations between nucleons are strongest for the proton-neutron system with the neutron given by $|d[ud]\rangle$ and the proton $|u[ud]\rangle$, a conclusion strongly supporting $n-p$  SRC as the EMC effect mechanism.  Experimental results indicate that $n-p$ correlations are up to $\sim 20$ times stronger than $n-n$ or $p-p$ \cite{Subedi:2008zz,Duer:2018sby} for relative NN momenta of $300-650~\rm MeV/c$.

In the 3-valence quark internal configuration of baryons, formation of a $[ud]$ scalar diquark will not be restricted to $n-p$ pairs.  The quark flavors of both $n-n$ and $p-p$ pairs allow diquark induced short-range correlations to form.  Neutron-proton pairings are still favored because there are more $[ud]$ combinations possible from the $n-p$ system than from  $p-p$ or $n-n$, shown explicitly in the last section of this work.  The ratio of $[ud]$ combinations in $n-n$ or $p-p$ vs. $n-p$ when nucleons have a 3-valence quark internal configuration is readily distinguished from the mandatory $n-p$ correlations of the quark-diquark configuration of nucleons.

This argument can be used to probe the internal configuration of valence quarks in nucleons over the EMC effect experimental parameter space.  If scalar $[ud]$ diquark formation is the underlying QCD model of the distortion of nucleon structure functions, then any experimental result showing only  neutron-proton NN short-range interactions indicates a predominantly quark-diquark nucleon structure in the corresponding physical parameter space, e.g. for nucleon knockout experiments with missing energy of $< 650~\rm MeV$ \cite{Korover:2020lqf}.  Regions of parameter space which do not show overwhelming neutron-proton type NN interactions would indicate indirect evidence for a lesser weighting of the  2-body quark-diquark configuration inside the nucleon and an inclusion of the 3-body configuration of valence quarks, discussed in detail in Sec.\ref{wavefunction}.  Evidence for suppression of the 20:1 dominance of $n-p$ SRC at high missing momentum in nucleon knockout experiments has recently been observed  \cite{Schmidt:2020kcl,Korover:2020lqf}.  

The unexpected distortion of nucleon structure functions and the relative strengths of $n-p$, $n-n$ and $p-p$ short-range correlations should have a theoretical foundation at the QCD level.  Diquark formation across nucleons offers an underlying $\rm SU(3)_C$ explanation for isospin dependent SRC as well as predicting significant deviations from $n-p$ dominance in $\rm A\!=\!3$ nuclei.  The deviations depend upon the internal configuration of the valence quarks inside nucleons.  There are two possibilities at lowest order (i.e. for 3-quark Fock states in the nucleon wavefunction), a 3-valence quark configuration, $|qqq\rangle$, and a quark-diquark configuration, $|q[ud]\rangle$.

\section{Diquark Properties}
\label{sec:mass}
Diquark properties have been studied for many decades, first in order to understand baryonic properties and reactions \cite{Cahill:1987qr} and later to understand exotic hadrons in QCD \cite{Jaffe:2004ph}.  Diquark masses and diquark effects in hadronic physics have been estimated using lattice QCD \cite{Francis:2021vrr,Watanabe:2021nwe}, baryon spectroscopy approximations \cite{Jaffe:2004ph}, instanton-induced effects \cite{Cristoforetti:2004rr,Schafer:1996wv}, light-front wavefunction analytic calculations \cite{Shuryak:2022wtk}, Nambu-Jona-Lasinio (NJL) calculations \cite{Roessner:2007gha,Thorsson:1989fw} and phenomenological models fit to data \cite{Abbott:1979je,Szczekowski:1988pq}, to name some of the most prominent methods.  The current work relies upon baryon mass spectroscopy prediction methods \cite{1981AmJPh..49..954G,Karliner:2014gca} as well as QCD hyperfine structure formulae \cite{DeRujula:1975qlm} for new diquark property estimations.  It is important to note that diquark properties require the group theory transformation of two quarks in the fundamental representation of $\rm SU(3)_C$ (the color triplet $3_C$) into an anti-fundamental representation of $\rm SU(3)_C$ (the color anti-triplet $\bar{3}_C$) - this is the definition of a diquark.  It is the diquark that forms a bound state, not simply the gluon exchange between two quarks \cite{Peskin:2019iig}.

The masses and binding energies of diquarks formed from $u$ and $d$ quarks may be calculated using a simple but powerful model of baryon and meson masses with sub-hadronic hyperfine interactions  \cite{1981AmJPh..49..954G},

\bea 
M_{\rm baryon}=\sum_{i=1}^{3} m_{i}+a^b \sum_{i< j}\left(\bar{\sigma}_{i} \cdot \bar{\sigma}_{j}\right) / m_{i} m_{j},
\label{eq:baryon-mass}
\eea 
\bea 
M_{\rm meson}=m_{1}+m_{2}+a^m \left(\bar{\sigma}_{1} \cdot \bar{\sigma}_{2}\right) / m_{1} m_{2},
\eea 
with $a^b$ a free parameter that is fit to baryon data and $a^m$ a free parameter fit to meson data.

The effective masses of the light quarks in baryons are found using Eq.\ref{eq:baryon-mass} and fitting to measured baryon masses,
\bea 
m_{u}^{b}=m_{d}^{b} \equiv m_{q}^{b}=363 ~\rm MeV, ~~m_{s}^{b}=538 ~\rm MeV,
\label{eq:parameters}
\eea 
with average error $\Delta m =5~\rm MeV$ when compared to measured ground state baryon masses  \cite{1981AmJPh..49..954G,Karliner:2014gca}.  These are effective masses of quarks bound in hadrons \cite{DeRujula:1975qlm}, not Standard Model Lagrangian masses where lattice QCD calculations give average $u$ and $d$ quark masses in the $\overline{\rm MS}$ renormalization scheme evaluated at energy scale $2~\rm GeV$ as  $\overline{m}_{u,d}=(3.9 \pm 0.3) ~ \mathrm{MeV}$ \cite{Dominguez:2018azt} and $m_{s}=(92.47 \pm 0.69) ~ \mathrm{MeV}$ \cite{Bazavov:2018omf}.

For diquarks consisting of $u$ and $d$ quarks, diquark parameters are calculated using the ground state masses of the spin-$\frac{1}{2}~$ isospin-0 $\Lambda(1116)$ and the spin-$\frac{1}{2}~$ isospin-1 $\Sigma^{0}(1193)$ baryons \cite{10.1093/ptep/ptaa104}.  Both baryons have quark flavor content $uds$ but differ by the spin and isospin assignments of their $u$ and $d$ quarks. Their mass splittings imply that the $\Lambda(1116)$ baryon contains the spin-0 isospin-0 $[ud]$ diquark and $\Sigma^{0}(1193)$ the spin-1 isospin-1 $(ud)$ diquark.  

Using the parameters from Eq.$~$\ref{eq:parameters} and Table \ref{Table:2}, the scalar $[ud]$ diquark binding energy is found to be 
\bea 
{\rm B.E.}_{[ud]} = m_{u}^{b}+m_{d}^b+m_{s}^{b} - M_{\Lambda}   = 148 ~\pm 9 ~ \rm MeV.
\label{eq:be}
\eea 
where the spin coupling between the $[ud]$ diquark and the $s$ quark is zero due to the cancellation between $u-s$ spin coupling and $d-s$ spin coupling, and the uncertainty is obtained by combining the errors in quadrature for each term (\textit{i.e.}, the uncertainties on effective masses are $5~\rm MeV$ \cite{1981AmJPh..49..954G,Karliner:2014gca} and the uncertainty on $M_{\Lambda}$ is given in \ref{Table:2} and is effectively negligible).

The $[ud]$ diquark mass then follows with the same uncertainty analysis,
\bea
m_{[ud]} = m_{u}^{b}+m_{d}^b - {\rm B.E.}_{[ud]} = 578~\pm 11 ~~\rm MeV.
\label{eq:scalarmass}
\eea

In order to estimate the masses of the vector diquarks, Eq.$~$\ref{eq:scalarmass} cannot be used because of the spin-spin interaction between the valence $s$ quark and the spin-1 diquark constituents (the interactions which canceled to zero for the $[ud]$).  Explicitly, the sub-hadronic hyperfine energies are given by 
\bea 
\Delta E^{\mathrm{HFS}}=\left\{\begin{array}{ll}
-3 a / m_{u}^{2} & (\Lambda) \\
a / m_{u}^{2}-4 a / m_{u} m_{s} & (\Sigma^{0})
\end{array}\right.
\eea 
with free parameter $a$ fit to data, $a /\left(m_{q}^{b}\right)^{2}=50 ~\rm MeV$, and $q=u,d$ \cite{1981AmJPh..49..954G,Karliner:2014gca}.  The mass of the $(ud)$ diquark can be found by using the 2-body meson mass formula instead,
\bea 
M_{(ud)}=m_d+m_u+a\left(\bar{\sigma}_{1} \cdot \bar{\sigma}_{2}\right) / m_{1} m_{2} 
\eea 
where $\bar{\sigma}_{1} \cdot \bar{\sigma}_{2}=+1$ for $S=1$, giving $m_{(ud)} = 776 ~\rm MeV$.  The masses of $(uu)$ and $(dd)$ are the same as the $(ud)$ in the approximation of $m_u=m_d$.

Binding energies for vector diquarks cannot be calculated using Eq.~\ref{eq:be} (with the appropriate baryons) because of spin-spin interactions that occur between the spin-1 diquark constituents and the remaining valence quark.  More to the point, isolated vector diquarks do not have binding energies because aligned spins add mass to the system, in the case of the $\Sigma$ baryons $\sim 50~ \rm MeV$.  However, there can be spin interactions between the vector diquark and the valence quarks in either nucleon of the correlated pair that lower the energy of the NN system.  The combined binding energy of vector diquarks and valence $s$ quarks is given here as an example of this effect, but the assumption in this work is that the $S=I=1$ diquarks are not required to couple to other spins and therefore $\rm B.E._{(qq)}=0$.

The combined binding energy of the spin-1 isospin-1 $(ud)$ diquark and the $s$ quark is found to be half that of the scalar diquark,
\bea 
{\rm B.E.}_{(ud)s} = m_{u}^{b}+m_{d}^b+m_{s}^{b} - M_{\Sigma^{0}}   = 71 ~\pm 9 ~ \rm MeV.~
\eea 

The combined binding energies of the remaining isospin triplet spin-1 diquarks $(uu)$ and $(dd)$ with the valence $s$ quark are found using the  $\Sigma^{-}(1197)$ baryon, quark content $dds$ and the $\Sigma^{+}(1189)$ baryon, quark content $uus$,
\bea 
{\rm B.E.}_{(uu)s} =  m_{u}^{b}+m_{d}^b+m_{s}^{b}  - M_{\Sigma^{+}} = 75 ~\pm 9 ~ \rm MeV,
\eea 
and similarly for ${\rm B.E.}_{(dd)s} =  67~\pm 9 ~ \rm MeV$.

Diquark parameter results are listed in Table \ref{Table:1} with relevant baryon parameters listed in Table \ref{Table:2}.

\begin{table} 
 \caption{Diquark properties}
\begin{tabular}{S|S|S|S|S}  \toprule 
    {Diquark} & {Binding Energy (MeV)} & {Mass (MeV)} & {Isospin $I$} & {Spin $S$} \\ \midrule
    {$[ud]$}  & {$ 148 \pm 9$} & {$578 \pm 11$} & 0 & 0 \\ \midrule
    {$(ud)$}  & {$0$}  & {$776 \pm 11$} & 1  & 1    \\ 
    {$(uu)$}  & {$0$}   & {$776 \pm 11$} & 1  & 1  \\
    {$(dd)$} & {$0$}  & {$776 \pm 11$} & 1  & 1 \\   \bottomrule
\end{tabular}
\label{Table:1}
\\[3pt] {\it {Uncertainties calculated using average  light quark mass errors $\Delta m_{q}=5 ~\rm MeV$ \cite{Karliner:2014gca}}}
\end{table}

\begin{table} 
 \caption{Relevant $\rm SU(3)_C$ hyperfine structure baryons \cite{10.1093/ptep/ptaa104}}
\begin{tabular}{S|S|S|S}  \toprule 
    {Baryon} & {Diquark-Quark content} & {Mass (MeV)}  & {$I\left(J^{P}\right)$} \\ \midrule
    {$\Lambda$}  & {$[ud]s$} & {$1115.683 \pm 0.006$} & {$0\left(\frac{1}{2}^{+}\right)$}  \\ \midrule
    {$\Sigma^{+}$}  & {$(uu)s$}  & {$1189.37~ \pm 0.07$} & {$1\left(\frac{1}{2}^{+}\right)$}     \\ 
    {$\Sigma^{0}$}  & {$(ud) s$}   & {$1192.642 \pm 0.024$} & {$1\left(\frac{1}{2}^{+}\right)$}    \\
    {$\Sigma^{-}$} & {$(dd) s$}  & {$1197.449 \pm 0.030$} & {$1\left(\frac{1}{2}^{+}\right)$}    \\   \bottomrule
\end{tabular}
\label{Table:2}
\\[3pt] {\it {$I\left(J^{P}\right)$ denotes the usual isospin $I$, total spin $J$ and parity $P$ quantum numbers, all have $L\!=\!0$ therefore $J=S$}}
\end{table}

The conclusion of the diquark parameter calculations is that a pair of nucleons in close proximity such that their valence quarks sense the attractive $3_C \otimes 3_C \rightarrow \overline{3}_C$ $~\rm SU(3)_C$ channel can lower their energy significantly by forming a bound scalar $[ud]$ diquark.  In addition, the $[ud]$ diquark mass is $\sim 200$ MeV less than the spin-1 diquark masses and $[ud]$ diquark formation lowers the mass of the system.

An estimate of quark-quark separation distances in the scalar $[ud]$ diquark are now discussed and compared to the NN separation distance in short-range correlated nucleon pairs.  We begin with an estimate of the q-q separation distance that makes use of the binding energy calculated in Sec.\ref{sec:mass} to find the radius of the diquark via a non-relativistic calculation from light nuclei physics \cite{PhysRev.137.B672}.  The assumption here is that the effective masses and motions of the quarks in the diquark can be investigated in the non-relativistic limit, an old approximation \cite{Bjorken:1979hv} that gives a result that agrees remarkably well with current calculations.  This estimate must be improved upon using relativistic calculations, as will be shown.  The non-relativistic relationship between the binding energy of a composite object and the radius of the object is given in \cite{PhysRev.137.B672} as 

\bea 
R \equiv \left(\frac{\rm{ln}(2)}{2 \mu B}\right)^{1 / 2}
\eea
where $B$ is the binding energy between the two bodies, here given by $148 ~ \rm MeV$ for the scalar diquark, and $\mu$ is the reduced mass of the two body system, given by $\mu_{[ud]}\equiv \frac{m_u m_d}{m_u + m_d} \sim 181~ \rm MeV$.  This non-relativistic formula is derived by taking the full-width half-max distance $R$ of an exponentially decaying wavefunction for a particle bound in a square well potential. Using the values calculated in Sec.\ref{sec:mass}, the radius of the scalar $[ud]$ diquark is estimated to be

\bea 
R_{[ud]} \sim 0.6 \times 10^{-15} ~\rm m.
\label{eq:radius}
\eea 
The diquark separation distances must be compared to NN separation distances for SRC. 

Experimental values of the relative momenta between  two short-range correlated nucleons include $400 ~\rm MeV/c$ for short-range neutron-proton correlations in $^{12}\rm C$ as well as a range of $300-600 ~ \rm MeV/c$ found in earlier studies \cite{Korover:2020lqf}.  The transition to inclusion of $p-p$ short-range correlations occurs in some targets at $\sim 800 ~\rm MeV/c$ \cite{Korover:2020lqf}.  These momenta correspond to a range of NN separation distances from $d_{\rm NN} = 0.25~\rm fm$ for $\Delta p\sim 800 ~\rm MeV/c$ to $d_{\rm NN}=0.66~\rm fm$ for $\Delta p \sim 300~\rm MeV/c$ by simple natural unit conversion where $1~\rm fm$ corresponds to $200~\rm MeV$.  The estimates suggest that valence quarks in correlated nucleons are well within the calculated quark-quark separation distance of the diquark. 

All relative NN momenta measured in short-range correlations as of today are above the nuclear Fermi momentum.  It is proposed that a sufficient NN separation distance in order for diquark formation to occur is the distance corresponding to the Fermi momentum of the nucleus, $\rm k_{F}= 250 ~\rm MeV/c$.  This translates, by the same natural unit conversion, to a separation distance of $\rm d_F = 0.79 ~\rm fm$. Therefore, the phenomenologically-driven separation distance between nearest neighbor nucleons sufficient for diquark induced SRC to occur is proposed to be 
\bea 
d_{NN} \lesssim d_{F} = 0.79 ~\rm fm.
\eea 

\section{Phenomenology of Diquark-Formation Induced SRC}
The CLAS collaboration work found that the ``isophobic'' nature of SRC - defined as the dominance of neutron-proton SRC rather than neutron-neutron or proton-proton - holds up to relative momentum up to $800~\rm MeV/c$, at which point the nn and pp SRC increase in number \cite{CLAS:2019vsb,CLAS:2018xvc}.  Thus far a case has been made for diquark formation across nucleons as the underlying cause of the observed short-range correlations in nucleon-nucleon pairs.  The case for SRC as the cause of the EMC effect was first made in papers circa 1990 \cite{CiofidegliAtti:1991ae,1990PhRvC..41.1100C} and has been intensively studied since then by many others.  A 2011 paper renewed interest in the topic \cite{PhysRevLett.106.052301} and recent data mining projects at Jefferson Lab by the CLAS collaboration have made strong connections between SRC and the EMC effect \cite{CLAS:2019vsb}.

Diquark creation across an NN pair is a QCD-based model of the SRC solution to the EMC effect.  The diquark model relies upon the established SRC-EMC connection and upon the diquark binding energy that lowers the mass of the 2-quark system.  Quark momentum depends on quark mass, even if the functional dependence $p(m_q)$ is unknown, and therefore diquark formation is likely to modify quark momentum distribution functions.  However likely it may be, the diquark model does not yet contain direct calculations of the measured structure function distortions $F_2(x_B)$ for all nuclear targets, which is the requirement for a definitive solution of the EMC effect. $F_2$ calculations for the modification of quark momentum distribution functions from quarks with lower effective masses due to diquark formation is currently a work in progress.  However, there are other tests for the model due to the isospin and spin content of the lowest mass diquark, $[ud]$.  Isospin dependent predictions for NN flavors can be made, as will be shown in the following section.

\subsection{Isospin Dependent Predictions}
The diquark-formation based SRC model is based on the low mass $[ud]$ diquark and the $u$ and $d$ quark flavors have an effect on the nuclear isospin dependence of SRC.  In the quark-diquark approach to nucleon structure \cite{Lichtenberg:1967zz,Qin:2019hgk,Brodsky:2014yha,1992PhLB..286...29B}, formation of the energetically favored scalar $[ud]$ diquark requires the nucleons to have available valence quarks with total isospin $I=0$.  In this scenario, each nucleon already contains one $[ud]$ diquark and therefore neutron-proton interactions dominate due to the available valence quark flavors,
\begin{align}
p-n &: ~~|[ud] \mathbf{u}\rangle ~|[ud]~\mathbf{d}\rangle  \\
p-p &: ~~|[ud] \mathbf{u}\rangle ~ |[ud]~\mathbf{u}\rangle \\
n-n &: ~~|[ud] \mathbf{d}\rangle ~ |[ud]~\mathbf{d}\rangle,
\end{align}
in the NN systems. Neutron-proton correlations were labeled ``isophobic'' short-range correlations and were first discovered at Brookhaven National Laboratory~\cite{2003PhRvL..90d2301T,Piasetzky:2006ai}.  They have subsequently been extensively studied by the CLAS collaboration at Jefferson Laboratory \cite{Subedi1476,Duer:2018sby,2019PhRvL.122q2502D}.  For two nucleons with a 3-valence quark internal configuration, any nuclear isospin combination of NN pairs can form $[ud]$ diquarks.  In this case, neutron-proton correlations are favored by simple counting arguments; a factor of $\frac{5}{4}$ in favor of $n-p$ because there are a maximum of 5 possible $[ud]$ diquarks in the $n-p$ system as opposed to 4 in the $n-n$ (or $p-p$) systems.  Both cases predict a distortion of quark distributions inside nuclei but the strength of the ``isophobic'' nature of short-range correlations differs.

\subsubsection{Predictions for A=3 nuclei}
\label{counting}
For $\rm A\!=\!3$ nuclei with all nucleons in a quark-diquark internal configuration, only $n-p$ SRC can form within the diquark induced SRC model.  By inspection of the 9-quark flavor content of the tritium nucleus $\rm ^3H$,
\bea 
{\rm ^3H}: |p\rangle |n\rangle |n\rangle \propto |\mathbf{u}[ud]\rangle |[ud]\mathbf{d}\rangle |[ud]\mathbf{d}\rangle
\eea 
and the $\rm ^3He$ nucleus, 
\bea 
{\rm ^3He}: |p\rangle |p\rangle |n\rangle \propto |\mathbf{u}[ud]\rangle |\mathbf{u}[ud]\rangle |[ud]\mathbf{d}\rangle,
\eea 
the case is made. The ratio of the number of $n-n$ or $p-p$ to $n-p$ SRC in this case is zero.  In contrast, 3-valence quark nucleon structure modifies the isospin dependence of NN SRC less dramatically.

The number of possible diquark combinations in $\rm A\!=\!3$ nuclei with nucleons in the 3-valence quark configuration is found by simple counting arguments.  First, the 9 quarks of $\rm ^3 He$  with nucleon location indices are written as:
\bea 
\begin{array}{llll}
{\rm N_1}:~ p\supset & u_{11} & u_{12} & d_{13} \\
{\rm N_2}: ~p\supset & u_{21} & u_{22} & d_{23} \\
{\rm N_3}: ~n\supset & u_{31} & d_{32} & d_{33}
\end{array}
\eea 
where the first index of $q_{ij}$ labels which of the 3 nucleons the quark belongs to, and the second index indicates which of the 3 valence quarks it is.  Diquark induced SRC requires the first index of the quarks in the diquark to differ, $[u_{ij}d_{kl}]$ with $i\neq k$.  The 4 possible combinations from $p-p$ SRC are listed below.
\bea 
u_{11} d_{23} \quad u_{12} d_{23} \\
u_{21} d_{13} \quad u_{22} d_{13}
\eea 
Short-range correlations from $n-p$ pairs have 10 possible combinations,
\bea 
\begin{array}{ll}
u_{11} d_{32} & u_{12} d_{32} \\
u_{11} d_{33} & u_{12} d_{33} \\
u_{21} d_{32} & u_{22} d_{32} \\
u_{21} d_{33} & u_{22} d_{33} \\
u_{31} d_{13} & u_{31} d_{23}
\end{array}
\eea 
which gives the number of $p-p$ combinations to $n-p$ combinations in this case as $\frac{2}{5}$.

Combining these results yields the following inequality for the isospin dependence of NN SRC:
\bea 
{\rm ^3He}:~~0 \leq \frac{\mathcal{N}_{pp}}{\mathcal{N}_{np}} \leq \frac{2}{5}
\eea 
where $\mathcal{N}_{N\!N}$ is the number of SRC between the nucleon flavors in the subscript.

The same argument may be made for $\rm ^3H$ due to the quark-level isospin-0 interaction, to find
\bea 
{\rm ^3H}:~~0 \leq \frac{\mathcal{N}_{nn}}{\mathcal{N}_{np}} \leq \frac{2}{5}.
\eea

 The fraction of same-nucleon NN SRC is related to the coefficients of the lowest order Fock states for $\rm A\!=\!3$ nuclei, as will be discussed in Sec.\ref{wavefunction}. It is not possible to obtain ratios greater than $\frac{2}{5}$ with diquark formation across nucleons.  Finding such experimental values would rule the model out.

\subsubsection{A=3 Nuclear wavefunction implications}
\label{wavefunction}
Individual nucleon wavefunctions at lowest order are dominated by two Fock states with unknown coefficients; the 3 valence quark configuration and the quark-diquark configuration,
\bea 
|\rm N\rangle = \alpha |qqq\rangle + \beta |q[qq]\rangle,
\eea 
where square brackets indicate the spin-0 $[ud]$ diquark.  The full $\rm A\!=\!3$ nuclear wavefunction is given by 
\begin{equation}
\begin{split}
|\Psi_{\rm A=3}\rangle &\propto (\alpha |qqq\rangle + \beta |q[qq]\rangle)(\alpha |qqq\rangle + \beta |q[qq]\rangle)\\
& (\gamma |qqq\rangle + \delta |q[qq]\rangle)
\end{split}\label{eq:wf}
\end{equation}
where the proton and the neutron are allowed to have different weights for each valence quark configuration.  This expands out to 
\begin{equation}
\begin{split}
|\Psi_{\rm A=3}\rangle &\propto {\alpha}^{2} \gamma |qqq\rangle^{3}+2 \alpha \beta \gamma |qqq\rangle^{2} |q[qq]\rangle \\
& \alpha^{2} \delta |qqq\rangle^{2} |q[qq]\rangle + {\beta}^{2} \gamma |qqq\rangle |q[qq]\rangle^{2} +\\
& 2 \alpha \beta \delta |qqq\rangle |q[qq]\rangle^{2} + {\beta}^{2} \delta |q[qq]\rangle^{3},
\end{split}
\end{equation}
with mixed terms demonstrating that it is not straightforward to map the $\frac{\mathcal{N}_{pp}}{\mathcal{N}_{np}}$ ratio to precise coefficients for each nucleon's Fock states.  A perhaps reasonable simplification is to assume that the proton and the neutron have the same coefficients for their 2-body and 3-body valence states, i.e. to set $\gamma = \alpha$ and $\delta = \beta$ in Eq. \ref{eq:wf}.  In this case, the nuclear wavefunction reduces to 

\begin{equation}
\begin{split}
|\Psi_{\rm A=3}\rangle &\propto {\alpha}^{3} |qqq\rangle^{3}+3 \alpha^{2} \beta |qqq\rangle^{2} |q[qq]\rangle \\
& + 3 {\beta}^{2} \alpha |qqq\rangle |q[qq]\rangle^{2} + {\beta}^{3} |q[qq]\rangle^{3}.
\end{split}
\end{equation}
With this assumption, the limiting cases yield exact knowledge of the coefficients: Experimental $\rm ^3H$ values of $\frac{\mathcal{N}_{nn}}{\mathcal{N}_{np}} = \frac{2}{5}$ implies $\beta \!=\! 0$, and $\frac{\mathcal{N}_{nn}}{\mathcal{N}_{np}} = 0$ implies $\alpha \!=\! 0$.  Intermediate values for $\frac{\mathcal{N}_{nn}}{\mathcal{N}_{np}}$ imply mixed internal configurations for the nucleons.  The case is the same for the $\frac{\mathcal{N}_{pp}}{\mathcal{N}_{np}}$ ratios of $\rm ^3He$. However, an important caveat must be made: If NN overlap is strong enough to place valence quarks so close together that they may break up an existing diquark to form a new one, this argument for the extremal values of the ratio is not valid.

\subsection{Diquark Formation in $\rm ^2H$}
\label{sec:hh}
Diquark formation across the proton and neutron in $\rm ^2H$ is different from other nuclei because the NN system must remain a color singlet and the ground state deuteron is spin-1 due to spin-statistics restrictions on the $n-p$ nuclear wavefunction.  At the quark level, a 6-quark wavefunction does not allow a ground state $[ud]$ diquark to form across nucleons with an internal quark-diquark configuration due to Bose statistics constraints upon diquark exchange.  This can be seen by considering both $n$ and $p$ in a quark-diquark structure and building a 3-diquark wavefunction with individual diquark wavefunctions given by Eq.~\ref{eq:diquark}.  The combination of 3 such wavefunctions, each antisymmetric in color, spin and isospin while symmetric in space, has Fermi statistics upon diquark exchange and is therefore forbidden.

Diquark formation in this case cannot work even with a $(ud)$ vector diquark correlation between the $n-p$ system.  The triple diquark state  in this case contains only two identical scalar diquarks but they pick up a minus sign upon exchange in the full wavefunction,
\bea 
|\Psi_{\rm ^2H}\rangle \propto \epsilon^{abc} (ud)^a[ud]^b[ud]^c
\eea 
due to color indices $abc$, and diquark formation induced SRC are therefore forbidden in this scenario.

For nucleons in a 3-valence quark internal structure, both scalar and vector diquarks can form between the $uud-udd$ quarks of the deuteron, but the overall configuration must respect the $S=1$ ground state.  Thus for a single $[ud]$ to form the remaining quarks must combine into a spin-1 state, together with any gluon and orbital angular momentum contributions.  In this scenario, a reduction in the EMC effect is predicted because the maximum total number of scalar diquarks (different from the number of possible combinations that can create $[ud]$) that can form across the $n-p$ system is 2 due to spin-statistics constraints on the system.  This reduction is as compared to structure function distortions extrapolated from the EMC-SRC effect in higher nuclei (higher nuclei not divided through by the deuteron's structure function)~\cite{Malace:2014uea}. Experimental results from the BONuS experiment at Jefferson Lab do show structure function distortions in the deuteron \cite{PhysRevC.92.015211}, with a possible reduction in strength as defined above \cite{Higinbotham:2010ye}.  

We note that the effect of diquark formation upon nuclear structure functions $F_A(x)$, defined by $F_A(x)\equiv F_{2}(x_B)= \sum_{f} x_B ~e_{f}^{2}\left(~q_{f}(x_B)+~\overline{q}_{\bar{f}}(x_B)\right)$ where $x_B$ is the fraction of the nucleon momentum carried by the struck quark, $e_f^2$ is the charge of the quark of flavor $f$ and $q_{f}(x_B)$ the probability to find a quark of flavor $f$ in the momentum range $x_{B} \in[x, x+d x]$, can be very roughly estimated by the diquark binding energy calulations of $\sim 150 ~\rm MeV$.  Lowering the mass of the NN system by such a large amount - the binding energy is on QCD interaction mass scales, not nuclear interaction mass scales - changes the momentum distribution of the quarks, regardless of the exact relationship between mass and momentum.  This means that the probability of a quark having a given momentum, encoded in the parton distribution function $q_f(x_B)$, is depleted when that quark forms a diquark bond.  This suppresses $F_A(x)$ for quarks within the nuclear medium that form diquarks across nucleon-nucleon pairs.  A precise estimation of the diquark formation effect on $F_A(x)$ is a long-term work in progress.

\section{Conclusions}
\label{sec:conclu}
Diquark creation across nucleons is proposed as the underlying QCD physics of nucleon-nucleon short-range correlations.  In addition, scalar $[ud]$ diquark formation is proposed to be the dominant QCD-level physics responsible for short-range NN correlation model explanation of the EMC effect. The short-range quark-quark QCD potential is attractive in the $\overline{3}_C$ channel of $\rm SU(3)_{C}$ and forms a bound state. The binding energy of the scalar isospin singlet $[ud]$ diquark is $148 ~\pm 9 ~ \rm MeV$ making it a highly energetically favorable bond.  Nucleon-nucleon separation distances sufficient for diquark formation are phenomenologically estimated from the Fermi momentum scale, proposed to be $d_{\rm NN}\geq 0.79~\rm fm$ between neighboring nucleons.  

The strength of the EMC effect in this model increases with increasing $A$ as any two nucleons in close enough proximity to sense the attractive $\rm SU(3)_C$ quark-quark potential will form a color anti-triplet.  The EMC effect strength increase with $A$ has been measured \cite{Schmidt:2020kcl}.

Isospin dependent NN short-range correlation studies by the E12-11-112 experiment at Jefferson Lab have just been published \cite{Li:2022fhh}.  The experiment observed the number of $p-p$ SRC (and assuming symmetry between ``mirror'' nuclei, the number of $n-n$ SRC) to $n-p$ SRC in $\rm A\!=\!3$ and they fell within the range $0 \leq \frac{\mathcal{N}_{nn/pp}}{\mathcal{N}_{np}} \leq \frac{2}{5}$ required for the diquark formation model to be correct. The upper limit of $\frac{2}{5}$ is a hard cutoff for the diquark formation model; higher values would have ruled the model out.  \\

 \subsection*{Acknowledgements}
  I am grateful to Stan Brodsky and Marek Karliner for many helpful discussions.  Thanks are due to Douglas Higinbotham and John Arrington for clarifying discussions on experimental results.  I thank Or Hen for his helpful comments on an earlier version of the manuscript.  I am indebted to Antonia Frassino and Francesco Coradeschi for useful discussions on Sec.\ref{sec:mass}. 
 This work is partially supported by the LDRD program of Lawrence Berkeley National Laboratory, the U.S. Department of Energy, Office of Science, Office of Nuclear Physics, under contract number DE-AC02-05CH11231 within the framework of the TMD Topical Collaboration, with additional support from the EIC Center at Jefferson Laboratory.

\bibliography{diquarkSRC}
\bibliographystyle{unsrt}

\end{document}